\documentclass{cimento}

\usepackage{amssymb,multirow}
\newcommand{\Kbar}{\,\overline{\!K}{}}
\def\K0bar{\Kbar{}^0}

\title{$\epsilon_K$ in the Standard Model and the kaon phase conventions}
\author{F.~Sala
\thanks{fsala@lpthe.jussieu.fr}}
\instlist{\inst{}
LPTHE, CNRS, UMR 7589, 4 Place Jussieu, F-75252, Paris, France}

\PACSes{\PACSit{11.30.Er}{Charge conjugation, parity, time reversal, and other discrete symmetries}
\PACSit{12.15.Ff}{Quark and lepton masses and mixing}
\PACSit{13.25.Es}{Decays of K mesons}
}

\begin{document}

\maketitle
\begin{abstract}
The parameter $\epsilon_K$, that quantifies CP violation in kaon mixing, is the observable setting the strongest constraints on new physics with a generic flavour and CP structure. While its experimental uncertainty is at the half percent level, the theoretical one is at the level of 15\%.
One of the largest sources of the latter uncertainty is the poor perturbative behaviour of the short-distance contribution of the box diagram with two charm quarks.
In this proceeding, based on ref.~\cite{Ligeti:2016qpi}, I summarise how that contribution can be removed, from the imaginary part of the mixing amplitude, by a rephasing of the kaon fields.
A first outcome is a mild reduction of the total theoretical uncertainty of $\epsilon_K$: while this might look counterintuitive at first sight, if different ``pieces'' (\textit{i.e.} short- and long- distance) of an observable are computed with different techniques, then it is possible to choose a phase convention where the total uncertainty of that observable is optimised.
Moreover, it is worthy to discuss if and how this freedom of rephasing, which has been somehow overlooked in the past, can help in making progress in lattice QCD computations of immediate relevance for $\epsilon_K$. 
\end{abstract}

\section{Motivation and Introduction}


CP violation (CPV) as a fundamental property of Nature was first established in the system of $K^0$ and $\K0bar$ mesons, with the measurement $\epsilon_K = 2.3 \times 10^{-3}$ (with a $\sim 20\%$ uncertainty) performed in 1964 at the Brookhaven National Laboratory~\cite{Christenson:1964fg}. Since then, $\epsilon_K$ has played a key role in establishing the CKM picture of flavour and CP violation, and in determining the related Standard Model (SM) parameters with precision. Today, the experimental determination of $\epsilon_K$ has reached the precision of half a percent, while the theoretical one is still well above the 10\% level. 

The achievement of a better control of the latter is strongly desirable. 
Indeed, among all observables, $\epsilon_K$ probes New Physics (NP) at some of the highest energy scales.
This holds for both the case in which the NP flavour structure is generic (\textit{i.e.} all flavour and CP violating operators are allowed, with coefficients of order one), and the case in which it is ``CKM-like'' (\textit{i.e.} only the operators that are also present in the SM are allowed, and each of them with the same parametrical suppression as in the SM), see \textit{e.g.} ref. \cite{Silvestrini:2016} for the updated UTfit~\cite{Bona:2007vi} results.
If one goes beyond those effective field theory considerations, still $\epsilon_K$ puts some of the most severe constraints on explicit NP models, for recent studies see \textit{e.g.} ref. \cite{Barbieri:2014tja} for supersymmetry and ref. \cite{Panico:2016ull} for composite Higgs models.
Thus, in the present situation where no clear evidence for NP has emerged in flavour and CP violating observables, $\epsilon_K$ could possibly hide deviations at the level of its current precision. If, conversely, a deviation from the SM would first show up in other observables (as the present data of $B$ meson decays seem to suggest \cite{Aaij:2015yra,Lees:2012xj,Lees:2013uzd,Huschle:2015rga,Aaij:2014ora}), then the level of consistency of these deviations with the properties of $K$ meson decays would be important to understand their NP origin.

That said, what are the present limiting factors of the $\epsilon_K$ sensitivity to NP?
The first one consists in the relatively poor knowledge of some CKM parameters, most notably $A$ (or equivalently $|V_{cb}| = \lambda^2 A$, $\lambda$ being the Cabibbo angle) and $\bar{\eta}$. This knowledge is expected to be improved substantially by upcoming measurements at Belle II and LHCb \cite{Aushev:2010bq,Bediaga:2012py} that, hopefully, will also help solving the current tension between the exclusive and inclusive determinations of $|V_{ub}|$ and $|V_{cb}|$~\cite{Agashe:2014kda}.

The largest non-parametric contribution to the $\epsilon_K$ theoretical uncertainty comes from the computation of the QCD corrections to the box diagram with two charm quarks, $\eta_{cc}$. The convergence of its perturbation series, 1 (LO), 1.38 (NLO), 1.87 (NNLO) is worrisome, the NNLO contribution being larger than the NLO one. Besides computing the NNLO part, ref. \cite{Brod:2011ty} has taken into account this apparently bad convergence by associating a ``large'' error to the central value of 1.87, $\eta_{cc} = 1.87 \pm 0.76$.
The QCD perturbative corrections to the analogous box diagrams which involve at least a top quark, $\eta_{tt} = 0.5765(65)$~\cite{Buras:1990fn} and $\eta_{ct} = 0.496(47)$~\cite{Brod:2010mj}, appear instead to be better behaved.
The bad convergence properties of the $\eta_{cc}$ perturbative series have induced different groups to treat it in a different way. Indeed, while CKMfitter~\cite{Hocker:2001xe} uses $\eta_{cc}$ as quoted in ref.~\cite{Brod:2011ty}, UTfit~\cite{Ciuchini:2000de} sticks to its NLO calculation. This contributes to the visibly different $\epsilon_K$ regions in the plots of the two collaborations, and somehow underlines the need to improve in that respect.

How then to make progress on the $\epsilon_K$ SM prediction?
The inclusion of the charm quark on the lattice would, ideally, replace the badly behaved perturbative QCD computation with a more controlled one, see \textit{e.g.} ref.~\cite{Christ:2012se} for a recent discussion specific to $\epsilon_K$, and ref.~\cite{Bai:2014cva} for updates.
Waiting for the first fruits of this proposal, we notice that ref.~\cite{Christ:2012se} (see Appendix A) already mentions that a non-standard approach to the $\epsilon_K$ computation \footnote{The suggestion of ref.~\cite{Christ:2012se} amounts to work with the substitution $\lambda_c = -\lambda_u - \lambda_t$, instead of the one $\lambda_u = -\lambda_c - \lambda_t$, where $\lambda_q = V_{qd} V_{qs}^*$ and $V$ is the CKM matrix. This suggestion is not to be confused with our proposal, which regards the freedom to rephase the kaon fields.
} might be needed to ease the aforementioned task.
This adds to the motivation to explore other non-standard approaches to the $\epsilon_K$ computation.

In this proceeding, I review the proposal of ref.~\cite{Ligeti:2016qpi} to set the $\eta_{cc}$ contribution to the $K^0-\K0bar$ mass mixing amplitude, $M_{12}$, purely real. This is achieved by employing the freedom to rephase the kaon fields, and it removes $\eta_{cc}$ from the theoretical prediction of $\epsilon_K$.
While physics should of course not depend on the phase convention chosen for the fields, some dependence remains numerically in this case due to the different techniques employed to compute different ``pieces'' entering the $\epsilon_K$ prediction, namely the calculation of $M_{12}$ (consisting of a short- and a long- distance ``piece'') and that of the decay mixing amplitude $\Gamma_{12}$ (long-distance). The rest of this contribution is organised as follows: in sect.~\ref{epsK_review} I review the ``usual'' SM prediction of $\epsilon_K$, in sect.~\ref{epsK_rephasing} I summarise the impact of a phase redefinition of the kaon fields, and discuss its implications, in sect.~\ref{conclusions} I conclude.

\section{An express review of $\epsilon_K$}
\label{epsK_review}

\subsection{$\epsilon_K$ independently of the kaon phase convention}
The CPV parameter $\epsilon_K$ is extracted from the decays into two pions of the neutral kaons mass eigenstates, dubbed as $K_L$ (longer lifetime, heavier eigenstate) and $K_S$ (shorter lifetime, lighter eigenstate), which are mixtures of the states $|K^0 \rangle = |d\bar{s}\rangle$ and $|\K0bar \rangle = |\bar{d}s \rangle$. $K_L$ and $K_S$ are the eigenvectors of $M - i\Gamma/2$, governing the Schr\"odinger equation of the $K^0-\K0bar$ system, where the mass ($M$) and the decay ($\Gamma$) mixing matrices are $2\times2$ Hermitian matrices.
Their mass and width splittings are $\Delta m = m_L - m_S = 3.484(6) \times 10^{-12}$ MeV and $\Delta \Gamma = \Gamma_L - \Gamma_S = - 7.3382(33) \times 10^{-12}$ MeV $(\simeq - 2 \Delta m)$. (We refer to the PDG~\cite{Agashe:2014kda} for all the measured values of observables reported here and in the rest of this contribution.)

Let us introduce the amplitude ratios\footnote{
Notice that this definition depends on the arbitrary relative phase between the states $K_L$ and $K_S$, and so it is not physical. Indeed, it is the phase-convention-independent quantity $\eta_f \equiv \big(\langle f | \mathcal{H} | K_L\rangle/\langle f| \mathcal{H} | K_S \rangle\big)/\big(\langle K^0|K_S\rangle/\langle K^0|K_L \rangle\big)\,$
that is measured in the interference of $|K_L\rangle$ and $|K_S\rangle$ decays in regeneration experiments. In the following, we will never need a non-zero relative phase between $K_L$ and $K_S$, so that the two definitions will be equivalent for our purposes.}
$\eta_f \equiv \langle f | \mathcal{H} | K_L\rangle/\langle f| \mathcal{H} | K_S \rangle \equiv A(K_L \to f)/ A(K_S \to f)$, in terms of which $\epsilon_K$ is defined as
\begin{equation}
\epsilon_K = \frac{2 \eta_{+-} + \eta_{00}}{3} \,,
\label{epsK_exp_def}
\end{equation}
for $f=\pi^+\pi^-$ and $\pi^0\pi^0$. Measurements yield $|\epsilon_K| = (2.228 \pm 0.011) \times 10^{-3}$, with a phase $\phi_\epsilon = (43.51\pm 0.05)^\circ$. One can  show that
\begin{equation}
{\rm Re}(\epsilon_K)  = \frac{{\rm Im}(M_{12}^* \Gamma_{12})}{4|M_{12}|^2 + |\Gamma_{12}|^2}\,
\label{Re_epsK}
\end{equation}
is valid up to relative orders $|\epsilon_K|^2$ and $|\omega| \epsilon'/\epsilon_K$, where $\omega = \langle (\pi\pi)_{I=2} | \mathcal{H} | K_S\rangle/\langle (\pi\pi)_{I=0} | \mathcal{H} | K_S \rangle \simeq 1/22$, and $\epsilon' = \omega (\eta_2 - \eta_0)/\sqrt{2} \simeq 3.7\times10^{-6}$.
Concerning the phase of $\epsilon_K$, the expression
\begin{equation}
\phi_\epsilon
\simeq \arctan \frac{2|M_{12}|}{|\Gamma_{12}|}
\label{phi_epsK}
\end{equation}
holds up to relative orders $|\omega^2 \epsilon'/\epsilon_K|$ and ${\rm Re}(\epsilon_K)^2$, and up to ratios of amplitudes that do not exceed a relative contribution of $10^{-2}$ to $\phi_\epsilon$.
The quantity $\arctan(-2 \Delta m/\Delta \Gamma) = 43.52^\circ$ is often referred to as ``superweak phase'', and differs from the value of $\phi_{\epsilon}$ measured at experiments by one part in $10^4$, so that the error of eq.~(\ref{phi_epsK}) neither exceeds that level.
Using eq.~(\ref{Re_epsK}) for Re$(\epsilon_K)$ and eq.~(\ref{phi_epsK}) for $\phi_\epsilon$ one obtains
\begin{equation}
\label{epsK_beauty}
\epsilon_K = \frac{e^{i\phi_\epsilon} \sin\phi_\epsilon}2
  \arg \bigg(\! - \frac{M_{12}}{\Gamma_{12}} \bigg)
= e^{i\phi_\epsilon} \cos\phi_\epsilon\, {\rm Im}(-M_{12}/\Gamma_{12})\,.
\end{equation}
The information about kaon mixing is exhausted by eq.~(\ref{epsK_beauty}) and by the following relations
\begin{equation}
\Delta m = 2|M_{12}| \,, \qquad \Delta \Gamma = - 2 |\Gamma_{12}| \,,
\label{deltaMG}
\end{equation}
which are valid up to relative orders $|\epsilon_K|^2$. The relative phase between $M_{12}$ and $\Gamma_{12}$ is $\pi + {\cal O}(|\epsilon_K|)$, because eq.~(\ref{Re_epsK}) implies that its sine is small, and  the eigenvalue equation $ 4 \,{\rm Re} (M_{12}^*\Gamma_{12}) = \Delta m \,\Delta \Gamma <0$ implies that its cosine is negative.

\subsection{$\epsilon_K$ in the usual phase convention}
To connect the phase convention independent expressions in eq.~(\ref{epsK_beauty}) to actual calculations, we need to consider how $M_{12}$ and $\Gamma_{12}$ are computed, namely
\begin{equation}
M_{12} = \frac{1}{2 m_K}\langle K^0|\mathcal{H}|\K0bar \rangle\,, \qquad
\Gamma_{12} = \sum_f A^*(K^0 \to f) A(\K0bar \to f)\,,
\label{def_M12}
\end{equation}
where $f$ denotes common final states of $K^0$ and $\K0bar$ decays.

The width mixing, $\Gamma_{12}$, is dominated by\footnote{
Notice that the minus sign in front of $|A_0^2|$ comes from the choice $\theta = \pi$ in $CP |K^0\rangle = e^{i\theta} |\K0bar\rangle$, $CP |\K0bar\rangle = e^{-i\theta} |K^0\rangle$. Further notice that the phase $\theta$ is not to be confused with the arbitrary phase of the kaon fields. It is on the latter that we base our rephasing.
}
$A_0^* \bar A_0 = - |A_0|^2\, e^{-2i\phi_0}$, where $A_I = A(K^0 \to (\pi\pi)_I)$ with ``$I$'' isospin eigenstate of the pions, and $\phi_0$ is the weak phase of the isospin-zero amplitude. $\phi_0$ depends on the arbitrary phase of the kaon fields, so that the quantity $\xi \equiv \tan\phi_0$ can take any value between $-\infty$ and $+\infty$. In phase conventions in which $|\xi| \ll 1$ one has 
\begin{equation}
 - \frac12 \arg (-\Gamma_{12}) = \xi = \phi_0
\end{equation}
up to relative orders $|\xi|^2$. The usual phase convention for the kaons indeed realises $|\xi| < |\epsilon_K| \ll 1$, and in that phase convention $\xi$ is determined either by the recent lattice computation Im$(|A_0|e^{i \phi_0}) = 1.90(1.22)(1.04)\times 10^{-11}$~GeV~\cite{Bai:2015nea}, or by the more precise lattice computation  Im$(|A_2|e^{i \phi_2}) = - 6.99(0.20)(0.84) \times 10^{-13}$ GeV~\cite{Blum:2015ywa} combined with the measured value of $|\epsilon'/\epsilon| = (1.66 \pm 0.23) \times 10^{-3}$ (therefore assuming the NP contribution to $\epsilon'/\epsilon$ is negligible).
The additional input of both determinations of $\xi$ is $|\omega| = |A_2/A_0|(1 + O(|\epsilon_K|)) \simeq 1/22$. Both determinations are thus phase-convention-dependent, and they result in the values for $\xi$ reported in table~\ref{tab:epsK_total}.

The choice of a phase convention where $|\xi| < |\epsilon_K|$ implies that $\{ \arg M_{12}\,,\, \arg\Gamma_{12} \} \lesssim O(|\epsilon_K|) \ll 1$ (mod $\pi$), so that one can write (using also that $\Delta m = 2|M_{12}|$)
\begin{equation}
\arg(-M_{12}/\Gamma_{12}) = \arg(M_{12}) - \arg(-\Gamma_{12})
  \simeq \frac{ 2 {\rm Im}M_{12}}{ \Delta m} + 2\xi \,.
\end{equation}
Eq.~(\ref{epsK_beauty}) for $\epsilon_K$ then becomes
\begin{equation}
\label{epsK_usual}
\epsilon_K = e^{i\phi_\epsilon} \sin\phi_\epsilon \bigg( \frac{{\rm Im}M_{12}}{\Delta m} + \xi \bigg) 
= e^{i\phi_\epsilon} \sin\phi_\epsilon  \bigg( \frac{{\rm Im}M_{12}^{\rm SD}}{\Delta m} + \xi    + \frac{{\rm Im}M_{12}^{\rm LD}}{\Delta m}\bigg)\,,
\end{equation}
where we have separated the short- and long- distance parts of the $M_{12}$. The calculation of Im$M_{12}^{\rm SD}$ relies on that of the perturbative SM box diagrams and their corrections, and on the one of the matrix element of the four-quark operator $\langle K^0| (\bar d_L\gamma_\mu s_L) (\bar d_L\gamma^\mu s_L) |\K0bar\rangle  = + \frac 23\, B_K(\mu) f_K^2 m_K^2$, where $B_K(\mu)$ expresses the deviation from the result of the vacuum insertion approximation.
One further defines $\widehat B_K$, to remove the $\mu$-dependence of $B_K(\mu)$. Concerning the long-distance part $M_{12}^{\rm LD}$, the most precise estimate has been given in ref.~\cite{Buras:2010pza} using chiral perturbation theory. It is conveniently expressed via the parameter
\begin{equation}
\rho = 1 + \frac{1}{\xi}\, \frac{{\rm Im}(M_{12}^{\rm LD})}{\Delta m} = 0.6 \pm 0.3\,.
\label{rho_def}
\end{equation}
Indeed, to separate the short- and long-distance pieces, the contribution of $M_{12}^{\rm LD}$ and $\xi$ is usually encoded in a factor multiplying $M_{12}^{\rm SD}$,
\begin{equation}
\label{def_kappa}
\kappa_\epsilon = \sqrt2\, \sin\phi_\epsilon
  \bigg( 1 + \rho\, \frac{\xi}{\sqrt2\, |\epsilon_K|} \bigg)\,.
\end{equation}

The above discussion yields the ``usual'' expression for $\epsilon_K$
\begin{equation}
\label{eps_old}
\epsilon_K =  \kappa_\epsilon\, e^{i\phi_\epsilon}\,\widehat{C}_\epsilon\,  |V_{cb}|^2 \lambda^2\,
  \bar{\eta}\, \Big\{ |V_{cb}|^2 (1-\bar{\rho}) \eta_{tt} S_0(x_t) 
  + \eta_{ct} S_0(x_t,x_c)
  - \eta_{cc} x_c \Big\},
\end{equation}
where\footnote{The NLO order in $\lambda$ is accidentally very suppressed, and the NNLO one is ${\cal O}(\lambda^{14})$.
}
 $x_q = [\overline m_q(\overline m_q) / m_W]^2$, the Inami-Lim functions $S_0$ can be found for example in ref.~\cite{Brod:2010mj}, and
$\widehat{C}_\epsilon =  f_K^2 \widehat{B}_K \, m_K\, m_W^2 \,G_F^2/(6 \sqrt2\,\pi^2\,\Delta m)\, = (2.806 \pm 0.049) \times 10^4\,.$
The reader interested in more details about the contents of this section is referred to section II of ref.~\cite{Ligeti:2016qpi} and references therein.

\section{$\epsilon_K$ without $\eta_{cc}$}
\label{epsK_rephasing}

\subsection{Rephasing the kaon fields}
With respect to the ``standard'' phase convention that lead to
Eq.~(\ref{eps_old}), one can rephase the kaon fields as
\begin{equation}
\label{kaons_rephase}
|K^0\rangle \to |K^0\rangle' = e^{i \lambda_c/|\lambda_c|} |K^0\rangle, \qquad
|\K0bar\rangle \to |\K0bar\rangle' = e^{-i \lambda_c/|\lambda_c|} |\K0bar\rangle\,,
\end{equation}
so that
\begin{eqnarray}
\label{rephase_M}
{\rm Im} M_{12} &\to & {\rm Im} M_{12}' = {\rm Im} M_{12} 
  \frac{{\rm Re} \lambda_c^2}{|\lambda_c^2|} +
  {\rm Re} M_{12} \frac{{\rm Im} \lambda_c^2}{|\lambda_c^2|} 
  \simeq {\rm Im} M_{12} + 2\lambda^4 A^2\bar{\eta}\, {\rm Re} M_{12} \,,
\\
\label{rephase_xi}
\xi &\to& \xi' = - \frac12 \frac{{\rm Im}(\Gamma_{12} \lambda_c^2)}
  {{\rm Re}(\Gamma_{12} \lambda_c^2)} 
\simeq - \frac{1}{2}\, \bigg( \frac{{\rm Im}\Gamma_{12}}{{\rm Re}\Gamma_{12}}
  + \frac{{\rm Im \lambda_c^2}}{{\rm Re \lambda_c^2}}\bigg)
  \simeq \xi - \lambda^4 A^2 \bar{\eta}\,.
\end{eqnarray}
%
%
Since $|{\rm Im}(\lambda_c)/{\rm Re}(\lambda_c)| < 10^{-3}$, this rephasing has a negligible impact on Re$(M_{12})$ (and thus on the expression for $\Delta m$), as opposed to the significant impact it has on Im$(M_{12})$ (and thus on the expression for $\epsilon_K$). This statement relies on the fact that, in every step, the phase-dependent errors never exceed a relative amount of $O(|\xi|^2)$, and in the new phase convention $|\xi'| < 10^{-3}$ still holds (see table~\ref{tab:epsK_total}).
Eqs.~(\ref{rephase_M}) and (\ref{rephase_xi}) do not know the way Im$(M_{12})$ and $\xi$ are computed, and indeed when they are plugged into eq. (\ref{epsK_usual}) for $\epsilon_K$, one sees that the phase convention dependence manifestly cancels out. However, if one brings forward the phase convention separately in $M_{12}^{\rm SD}$, in $\xi$ and in $M_{12}^{\rm LD}$, then
\begin{itemize}
\item[$\diamond$] in $M_{12}^{\rm SD}$, the $\eta_{cc}$ piece becomes purely real (indeed, it was proportional to the phase we rotated away), and thus it does not manifestly contribute to $\epsilon_K$;
\item[$\diamond$] the $\xi$ value and error change according to eq.~(\ref{rephase_xi}), see table~\ref{tab:epsK_total};
\item[$\diamond$] using eqs.~(\ref{rephase_M}) and (\ref{rephase_xi}), and that $|\xi| \ll |\xi'|$, we obtain\footnote{Notice that we do not perform the estimate of ref.~\cite{Buras:2010pza}, that lead to eq.~(\ref{rho_def}), directly in the new phase convention. We just use that result in the usual phase convention, and rephase it. 
}
\begin{equation}
\label{def_kappa_prime}
\rho' = 0.6 \pm 0.2\,, \quad {\rm where} \quad \kappa_\epsilon' = \sqrt{2}\sin\phi_\epsilon \times 
  \left(1 + \rho' \frac{\xi'}{\sqrt{2} |\epsilon_K|} \right).
\end{equation}
The expression of $\epsilon_K$ in our phase convention then becomes
\begin{equation}
\epsilon_K =  \kappa_\epsilon'\, e^{i\phi_\epsilon} \,\widehat{C}_\epsilon\,  |V_{cb}|^2 \lambda^2\,
  \bar{\eta}\, \Big\{ |V_{cb}|^2 (1-\bar{\rho}) \eta_{tt} S_0(x_t) 
  + \eta_{ct} S_0(x_t,x_c) \Big\}.
\label{epsK_readytouse}
\end{equation}

\end{itemize}

\subsection{Implications of the rephasing}
\begin{table}[t]
\begin{tabular}{cc|c|cc}
\hline
& CKM inputs  &$|\epsilon_K| \times 10^3$ & 
  \raisebox{0pt}[14pt][0pt]{$\kappa_\epsilon^{(\prime)}$}  & 
  $\xi^{(\prime)} \times 10^4$\\
\hline
\multirow{2}{*}{usual evaluation}
&  tree-level   &
$ 2.30\pm 0.42$  &  $0.963 \pm 0.010$  & $-0.57 \pm 0.48$\\
&  SM CKM fit  & 
$2.16 \pm 0.22$  &  $0.943 \pm 0.016$  & $-1.65 \pm 0.17$\\
\hline
\multirow{2}{*}{our evaluation}
&  tree-level  &
$2.38 \pm 0.37$  &  $0.844 \pm 0.044$  & $-6.99 \pm 0.92$\\
&  SM CKM fit 
& $2.24 \pm 0.19 $  &  $0.829 \pm 0.049$  & $-7.83 \pm 0.26$\\
\hline
\end{tabular}
\caption{Present value of $\epsilon_K$ in the usual evaluation (upper part) and in our evaluation (lower part). We also show the values of $\kappa_\epsilon$ and $\xi$ in the upper part, and $\kappa_\epsilon'$ and $\xi'$ in the lower part.}
\label{tab:epsK_total}
\end{table}

\begingroup

\begin{table}[t]
\begin{tabular}{cc|ccc|ccc|c}
\hline
&CKM inputs  &  $\eta_{cc}$ & $\eta_{ct}$  &  $\kappa_\epsilon^{(\prime)}$ 
  &  $|V_{cb}|$ &  $\bar{\eta}$  &  $\bar{\rho}$
  &  $|{\Delta \epsilon_K}/{\epsilon_K}|_{\rm tot.}$\\
\hline
\multirow{2}{*}{usual evaluation}
&  tree-level  & 7.3\%& 4.0\%& 1.1\%
  &11.1\%  &10.4\%  & 5.4\%  & 18.4\%\\
&  SM CKM fit   & 7.4\%  & 4.0\%  &  1.7\% 
  & 4.2\%  &  2.0\%  &  0.8\%  & 10.2\%\\
\hline
\multirow{2}{*}{our evaluation}
&  tree-level      &  ---  &  3.4\%  &  5.2\%  &  9.5\%  &  8.9\%  &  4.5\%  & 15.6\%\\
&  SM CKM fit   &  ---  & 3.4\%  &  5.9\%  & 3.6\%  &  1.7\%  &  0.7\%  & 8.4\%\\
\hline
\end{tabular}
\caption{The present error budget of $\epsilon_K$ in the usual evaluation (upper part) and using our evaluation (lower part).  The parameters with a corresponding uncertainty above 2\% are shown.}
\label{tab:error_budget}
\end{table}
\endgroup

We refer to two sets of input CKM parameters, one from the SM CKM fit results (that therefore assumes that the SM determines all observables), and one from the tree-level observables only fit (applicable even if TeV-scale new physics affects the loop-mediated processes), which bares of course larger uncertainties. See ref.~\cite{Ligeti:2016qpi} for the values of the CKM parameters $\lambda, A, \bar{\eta}, \bar{\rho}$ in the two cases.

The partial and total error budgets of $\epsilon_K$ are reported in table~\ref{tab:error_budget}, in both the new and the usual phase conventions.  One sees there that the convention we propose yields to a mild gain in the total $\epsilon_K$ uncertainty, and increases the relative importance of the long-distance contribution, encoded in $\kappa_\epsilon'$. Its relative importance will increase in the future, since Belle II and LHCb are expected to improve on the CKM inputs, bringing their tree-level determinations to a precision comparable to (or better than) the one of the current SM CKM fit~\cite{Charles:2013aka}, see table~\ref{tab:error_budget}.
One might then argue that the LD estimate of ref.~\cite{Buras:2010pza} should be rediscussed, strengthening the need to make progress on the lattice QCD determination of Im($M_{12}^{\rm LD}$). A clarification of the roles of the different unphysical phases can help in this direction.



\section{Conclusions}
\label{conclusions}
Without any clear deviation from the SM picture of flavour and CP violation, it is hard, if not impossible, to obtain clues about a more fundamental theory of flavour.
Among all observables, $\epsilon_K$ is sensitive to some of the highest energies, and sets some of the strongest constraints on explicit NP models. Therefore, it is crucial to improve its theoretical determination, whose uncertainty is much larger than the experimental one.

In this conference proceeding, I explained the recent proposal of ref.~\cite{Ligeti:2016qpi} to remove from $\epsilon_K$ the largest source of non-parametric error, $\eta_{cc}$, via a rephasing of the kaon fields. As a consequence, the total uncertainty of the $\epsilon_K$ SM prediction is slightly reduced, and the source of the largest (after some of the CKM inputs) error comes from the long distance contribution to $M_{12}$.
Our discussion also helps clarifying the role of different phases in the computation, and we think it would be important to understand if it can help achieving a better lattice control of the above long-distance contribution.


\acknowledgments
I thank the organisers of ``Les Rencontres de physique de la Vall\'ee d'Aoste''  for the possibility to present this contribution, as well as for the nice and stimulating atmosphere they created in La Thuile.
I thank Zoltan Ligeti for the pleasant collaboration on the subject of this proceeding, and for comments on the manuscript.
I am supported by the European Research Council ({\sc Erc}) under the EU Seventh Framework Programme (FP7/2007-2013)/{\sc Erc} Starting Grant (agreement n.\ 278234 --- `{\sc NewDark}' project).


\begin{thebibliography}{0}
\bibitem{Ligeti:2016qpi}
  Z.~Ligeti and F.~Sala,
  \TITLE{A new look at the theory uncertainty of epsilon\_K},
  arXiv:1602.08494 [hep-ph].


\bibitem{Christenson:1964fg}
  J.~H.~Christenson, J.~W.~Cronin, V.~L.~Fitch and R.~Turlay,
  \TITLE{Evidence for the 2 pi Decay of the k(2)0 Meson},
  Phys.\ Rev.\ Lett.\  {\bf 13} (1964) 138.
  
  \bibitem{Silvestrini:2016}
  L.~Silvestrini [UTfit Collaboration],
  \TITLE{Model-independent constraints on $\Delta F=2$ operators and the scale of new physics},
  talk at ``pp\@LHC 2016'', Pisa University, 17 May 2016.

  
 \bibitem{Bona:2007vi}
  M.~Bona {\it et al.} [UTfit Collaboration],
  \TITLE{Model-independent constraints on $\Delta F=2$ operators and the scale of new physics},
  JHEP {\bf 0803} (2008) 049
  [arXiv:0707.0636 [hep-ph]].
  
  
  \bibitem{Barbieri:2014tja}
  R.~Barbieri, D.~Buttazzo, F.~Sala and D.~M.~Straub,
  \TITLE{Flavour physics and flavour symmetries after the first LHC phase},
  JHEP {\bf 1405} (2014) 105
  [arXiv:1402.6677 [hep-ph]].
    
  \bibitem{Panico:2016ull}
  G.~Panico and A.~Pomarol,
  \TITLE{Flavor hierarchies from dynamical scales},
  arXiv:1603.06609 [hep-ph].
  
  
  \bibitem{Aaij:2015yra}
  R.~Aaij {\it et al.} [LHCb Collaboration],
  \TITLE{Measurement of the ratio of branching fractions $\mathcal{B}(\bar{B}^0 \to D^{*+}\tau^{-}\bar{\nu}_{\tau})/\mathcal{B}(\bar{B}^0 \to D^{*+}\mu^{-}\bar{\nu}_{\mu})$},
  Phys.\ Rev.\ Lett.\  {\bf 115} (2015) no.11,  111803
   Addendum: [Phys.\ Rev.\ Lett.\  {\bf 115} (2015) no.15,  159901]
  [arXiv:1506.08614 [hep-ex]].
  
  \bibitem{Lees:2012xj}
  J.~P.~Lees {\it et al.} [BaBar Collaboration],
  \TITLE{Evidence for an excess of $\bar{B} \to D^{(*)} \tau^-\bar{\nu}_\tau$ decays},
  Phys.\ Rev.\ Lett.\  {\bf 109} (2012) 101802
  [arXiv:1205.5442 [hep-ex]].
  
  \bibitem{Lees:2013uzd}
  J.~P.~Lees {\it et al.} [BaBar Collaboration],
  \TITLE{Measurement of an Excess of $\bar{B} \to D^{(*)}\tau^- \bar{\nu}_\tau$ Decays and Implications for Charged Higgs Bosons},
  Phys.\ Rev.\ D {\bf 88} (2013) no.7,  072012
  [arXiv:1303.0571 [hep-ex]].
  
  \bibitem{Huschle:2015rga}
  M.~Huschle {\it et al.} [Belle Collaboration],
  \TITLE{Measurement of the branching ratio of $\bar{B} \to D^{(\ast)} \tau^- \bar{\nu}_\tau$ relative to $\bar{B} \to D^{(\ast)} \ell^- \bar{\nu}_\ell$ decays with hadronic tagging at Belle},
  Phys.\ Rev.\ D {\bf 92} (2015) no.7,  072014
  [arXiv:1507.03233 [hep-ex]].
  
  \bibitem{Aaij:2014ora}
  R.~Aaij {\it et al.} [LHCb Collaboration],
  \TITLE{Test of lepton universality using $B^{+}\rightarrow K^{+}\ell^{+}\ell^{-}$ decays},
  Phys.\ Rev.\ Lett.\  {\bf 113} (2014) 151601
  [arXiv:1406.6482 [hep-ex]].
  
  
  




\bibitem{Aushev:2010bq}
  T.~Aushev {\it et al.},
  \TITLE{Physics at Super B Factory},
  arXiv:1002.5012 [hep-ex].
  
  \bibitem{Bediaga:2012py}
  R.~Aaij {\it et al.} [LHCb Collaboration],
  \TITLE{Implications of LHCb measurements and future prospects},
  Eur.\ Phys.\ J.\ C {\bf 73} (2013) no.4,  2373
  [arXiv:1208.3355 [hep-ex]].
  



\bibitem{Agashe:2014kda} 
  K.~A.~Olive {\it et al.}  [Particle Data Group Collaboration],
  Chin.\ Phys.\ C {\bf 38}, 090001 (2014).


\bibitem{Brod:2011ty}
  J.~Brod and M.~Gorbahn,
  \TITLE{Next-to-Next-to-Leading-Order Charm-Quark Contribution to the CP Violation Parameter epsilon\_K and Delta M\_K},
  Phys.\ Rev.\ Lett.\  {\bf 108} (2012) 121801
  [arXiv:1108.2036 [hep-ph]].
  
  
  \bibitem{Buras:1990fn}
  A.~J.~Buras, M.~Jamin and P.~H.~Weisz,
  \TITLE{Leading and Next-to-leading {QCD} Corrections to $\epsilon$ Parameter and $B^0 - \bar{B}^0$ Mixing in the Presence of a Heavy Top Quark},
  Nucl.\ Phys.\ B {\bf 347} (1990) 491.
  
  \bibitem{Brod:2010mj}
  J.~Brod and M.~Gorbahn,
  \TITLE{Epsilon\_K at Next-to-Next-to-Leading Order: The Charm-Top-Quark Contribution},
  Phys.\ Rev.\ D {\bf 82} (2010) 094026
  [arXiv:1007.0684 [hep-ph]].
  
  
  \bibitem{Hocker:2001xe} 
  A.~H\"ocker, H.~Lacker, S.~Laplace and F.~Le Diberder,
\TITLE{A New approach to a global fit of the CKM matrix},
  Eur.\ Phys.\ J.\ C {\bf 21}, 225 (2001)
  [hep-ph/0104062];
  J.~Charles {\it et al.},
\TITLE{$CP$ violation and the CKM matrix: Assessing the impact of the asymmetric $B$ factories},
  Eur.\ Phys.\ J.\ C {\bf 41} (2005) 1
  [hep-ph/0406184];
and updates at http://ckmfitter.in2p3.fr/.

\bibitem{Ciuchini:2000de} 
  M.~Ciuchini {\it et al.},
  \TITLE{2000 CKM triangle analysis: A Critical review with updated experimental inputs and theoretical parameters},
  JHEP {\bf 0107}, 013 (2001)
  [hep-ph/0012308];
  M.~Bona {\it et al.} [UTfit Collaboration],
  \TITLE{The 2004 UTfit collaboration report on the status of the unitarity triangle in the standard model},
  JHEP {\bf 0507}, 028 (2005)
  [hep-ph/0501199];
and updates at http://utfit.org/.

  
\bibitem{Christ:2012se}
  N.~H.~Christ {\it et al.} [RBC and UKQCD Collaborations],
  \TITLE{Long distance contribution to the KL-KS mass difference},
  Phys.\ Rev.\ D {\bf 88} (2013) 014508
  [arXiv:1212.5931 [hep-lat]].
  
  \bibitem{Bai:2014cva}
  Z.~Bai, N.~H.~Christ, T.~Izubuchi, C.~T.~Sachrajda, A.~Soni and J.~Yu,
  \TITLE{$K_L-K_S$ Mass Difference from Lattice QCD},
  Phys.\ Rev.\ Lett.\  {\bf 113} (2014) 112003
  [arXiv:1406.0916 [hep-lat]];
  Z.~Bai,
  \TITLE{$K_L$ - $K_S$ mass difference computed with a 171 MeV pion mass},
  PoS LATTICE {\bf 2014} (2015) 368
  [arXiv:1411.3210 [hep-lat]].
  
  
 

  
  
    \bibitem{Bai:2015nea} 
  Z.~Bai {\it et al.} [RBC and UKQCD Collaborations],
  \TITLE{Standard Model Prediction for Direct CP Violation in $K \to \pi \pi$ Decay},
  Phys.\ Rev.\ Lett.\  {\bf 115}, no. 21, 212001 (2015)
  [arXiv:1505.07863 [hep-lat]].

\bibitem{Blum:2015ywa} 
  T.~Blum {\it et al.},
  \TITLE{$K \rightarrow \pi\pi$ $\Delta I=3/2$ decay amplitude in the continuum limit},
  Phys.\ Rev.\ D {\bf 91}, no. 7, 074502 (2015)
  [arXiv:1502.00263 [hep-lat]].
  
  \bibitem{Buras:2010pza} 
  A.~J.~Buras, D.~Guadagnoli and G.~Isidori,
  \TITLE{On $\epsilon_K$ Beyond Lowest Order in the Operator Product Expansion},
  Phys.\ Lett.\ B {\bf 688}, 309 (2010)
  [arXiv:1002.3612 [hep-ph]].
  
  \bibitem{Charles:2013aka}
  J.~Charles, S.~Descotes-Genon, Z.~Ligeti, S.~Monteil, M.~Papucci and K.~Trabelsi,
  \TITLE{Future sensitivity to new physics in $B_d, B_s$, and K mixings},
  Phys.\ Rev.\ D {\bf 89} (2014) no.3,  033016
  [arXiv:1309.2293 [hep-ph]].


  
   
  
  
  









\end{thebibliography}
\end{document}